\documentclass[12pt]{article}
\usepackage{setspace}
\usepackage{geometry}
\usepackage{multirow}
\usepackage{color}
\usepackage[affil-it]{authblk}

\usepackage{graphicx}
\graphicspath{{../../Figures/2013_14/}{../../screenshots/}{./Figures/}}

\usepackage{amsmath,amssymb}
\usepackage{caption}
\usepackage{subcaption}
\usepackage{booktabs}
\usepackage{float}
{\newgeometry{left=1in,right=1in,top=1in,bottom=1in}

\usepackage{ wasysym }
\usepackage[round]{natbib}

\begin{document}

\title{The Impact of Large Scale Promotions on the Sales and Ratings of Mobile Apps: Evidence from Apple's App Store}
\date{\today}
\author{Georgios Askalidis\footnote{gask@u.northwestern.edu}}
\affil{Northwestern University}
%\email{gask@u.northwestern.edu}
%\affiliation{Northwestern University}
\maketitle

%\doublespacing
%\title{\textbf{\large{The Impact of Large Scale Promotions on the Sales and Ratings of Mobile Apps: Evidence from Apple's App Store}}}
%\date{}
%\author{
%Georgios Askalidis\\
%\affaddr{Northwestern University}
%\email{gask@u.northwestern.edu}
%}
\newcommand{\vp}{{\varphi}}
\doublespacing
\maketitle
\begin{abstract}
We study four promotions offered on Apple's mobile app store that vary in user exposure, price discount and redemption procedure. We find that promotions that are full-price discounted and digital (i.e., the redemption procedure is a few taps on the smartphone) are the ones that cause the largest increase in downloads. Investigating the effect on ratings, we find neutral or positive effects on the ratings for all full-price discounted promotions but negative short-term effects on the ratings of apps promoted by substantial, but only partial, price discounts. Furthermore, we find that high barrier promotions induce a rich-get-richer effects, with apps that were popular before the promotion receiving, generally, larger benefits. In contrast, low barrier promotions cause, on average, the same increase in sales for all participating apps regardless their characteristics or their previous popularity. Finally, we explore the effect of these promotions on the competition of the featured apps and find negative externalities when the promotion is low barrier and positive when it's high barrier.
\end{abstract}
\newpage
%\category{H.4}{Information Systems Applications}{Miscellaneous}
%\category{D.2.8}{Software Engineering}{Metrics}[complexity measures, performance measures]

%\terms{Theory}

%\keywords{ACM proceedings, \LaTeX, text tagging} % NOT required for Proceedings

\section{Introduction}

\par Mobile apps have become an economy with a market size of \$25 Billion in 2013\footnote{Wall Street Journal (http://on.wsj.com/1Haufei)} and with a projected market size of \$77 Billion by 2017\footnote{entrepreneur.com (http://goo.gl/GdOa3P)}. As of mid 2016, The Google Play store for Android apps and iOS App Store for iOS apps, owned and operated by Google Inc. and Apple Inc. respectively, are the two largest mobile app stores, each with more than 1.5 Million mobile apps available to download\footnote{statista.com (http://goo.gl/dm1aw2)}.  As of July 2013 the two mobile app stores had seen more than 50 Billion app downloads\footnote{statista.com (http://goo.gl/knUMyp)} \footnote{statista.com (http://goo.gl/dcqFks)} each. Apple announced100 Billion downloads in June 2015. With such volumes, it's impossible for users to be aware even of a small fraction of all the available apps: the average iOS and Android smartphone user used less than 30 apps per month at the end of 2013\footnote{nielsen.com (http://goo.gl/QkSPCw)}.
\par This incentivizes the owners of the stores to segment their content into easily navigable segments such as `Editor's Choice',  `Essentials' or `New to the Store' to help their users find their most preferred apps. Similarly, creators of apps try to induce word of mouth and publicity for their apps, for example, via free `light' versions of their (not-free) app or by lowering the app's price for a limited time. 
\par But promotions could come with their own caveats. For example, a price reduction could attracts users that are not in the target group of an app and hence lead to suboptimal user experiences which, in turn, can lead to lower online reviews. These lower online reviews, can significantly impact the future performance of any product and service. On the other hand, developers usually strive to be featured in their stores, perhaps without having a complete understanding of the benefits but also risks that such promotions can have. Our paper is contributing towards a better understanding of the effects that such promotions can have on the sales and ratings of featured apps, as well their competition.
\par We focus on four such promotion campaigns in the iOS App Store. Two of them are versions of a larger promotion offered by the coffee chain Starbucks, called `iOS Pick of the Week'. With both promotions, Starbucks offers a one-time redemption code that customers can use to get a featured app for free. The difference between the two promotions is that one distributes the redemption codes via printed coupons in physical stores (we will refer to this distribution method as `In-Store') and the other from within the Starbucks iOS app (we will refer to this distribution method as `Digital'). The sets of apps featured in these two versions of the promotion are non overlapping. The other promotions we study were offered by Apple itself. 
The first, `App of the Week', is a recurring weekly free offering of an app that is generally not free otherwise. 
The other ones, are four one-time promotions that run for one week each the third week of December 2014, the third week of July 2015, the fourth week of May 2015 and the third week of December 2015. Three of the promotions were called `Amazing Apps \& Games for 99\cent each' and the fourth was called `Handpicked Apps \& Games for \$0.99'. The general theme was the same in all four promotions: Apple chose 30 apps (usually 15 of them were games) and offered them for \$0.99 each for a week. Since the last four promotions are so similar in spirit and execution we bundle them together and refer to them simple as the Apple Amazing promotion.
\par One of the key differences between the two promotions offered by Apple and the two promotions offered by Starbucks is that the two promotions offered by Apple feature large banners on the front page of the App Store, and the featured apps were sold for the reduced price for all users (even those who were unaware of the promotion before their visit). In contrast, Starbucks' promotions were valid only for customers that had acquired a redemption code either in store or though the iOS app. All other users would have to pay the full price to get the app.% See Figure \ref{screenshots} for some screenshots of the promotions.
\par Even though these promotions were offered both inside and outside of the United States, in this study we focus only on the US App Store and US Starbucks stores. 
\subsection{Main Insights}
\par We focus on two main characteristics of the promotions: steepness of price discount and ease of redemption. We utilize the similarities and differences between the four promotions in our study to understand the effect each of these two promotion characteristics had on the sales and ratings of the promoted apps. We find that digital promotions, which are easy to redeem, accompanied by a full price discount are the promotions that cause the largest increase in sales, while having neutral or even positive effects on the ratings. In-store coupons, which are slightly less trivial to redeem, cause a smaller increase in sales but also have no effect on the ratings. Non-full price discounted promotions cause the smallest (but still positive) increase in sales but can also have a negative effect on the ratings. 
\par Furthermore, we explore the effect of various app characteristics in the success of the app's promotion. We find that in digital and full-price discounted promotions, all featured apps receive on average the same increase in sales. But in promotions with higher barriers (such as less trivial redemption procedure or non-zero price) apps that were more popular before the promotion are the ones that will receive the largest increases in sales. This indicates that when the user is offered an app that is free and (literally) a few taps away, the specific characteristics of the app will not matter much on their decision to redeem the offer or not. But when the barriers are higher, users are more likely to do the effort (or pay the price) if it's a more established app that they are familiar with.
\par Finally, we are also interested in understanding the effect that these promotions had on the apps competing with the promoted apps. We find that full price discounted promotions to have a significant negative effect on the sales of their competition but not full price discounted promotions have a positive effect on the sales of their competition. We interpret this result as a sign that users are intrigued and interested when they are exposed to an app through a promotion. If the app is not free, the user will not commit immediately but instead will explore the alternatives. When the app is free, the user will have little incentive to search for alternatives.
\par All these insights can directly help practitioners better design their marketing campaigns.

\begin{figure}[h!]
\centering
\includegraphics[width=0.45\textwidth]{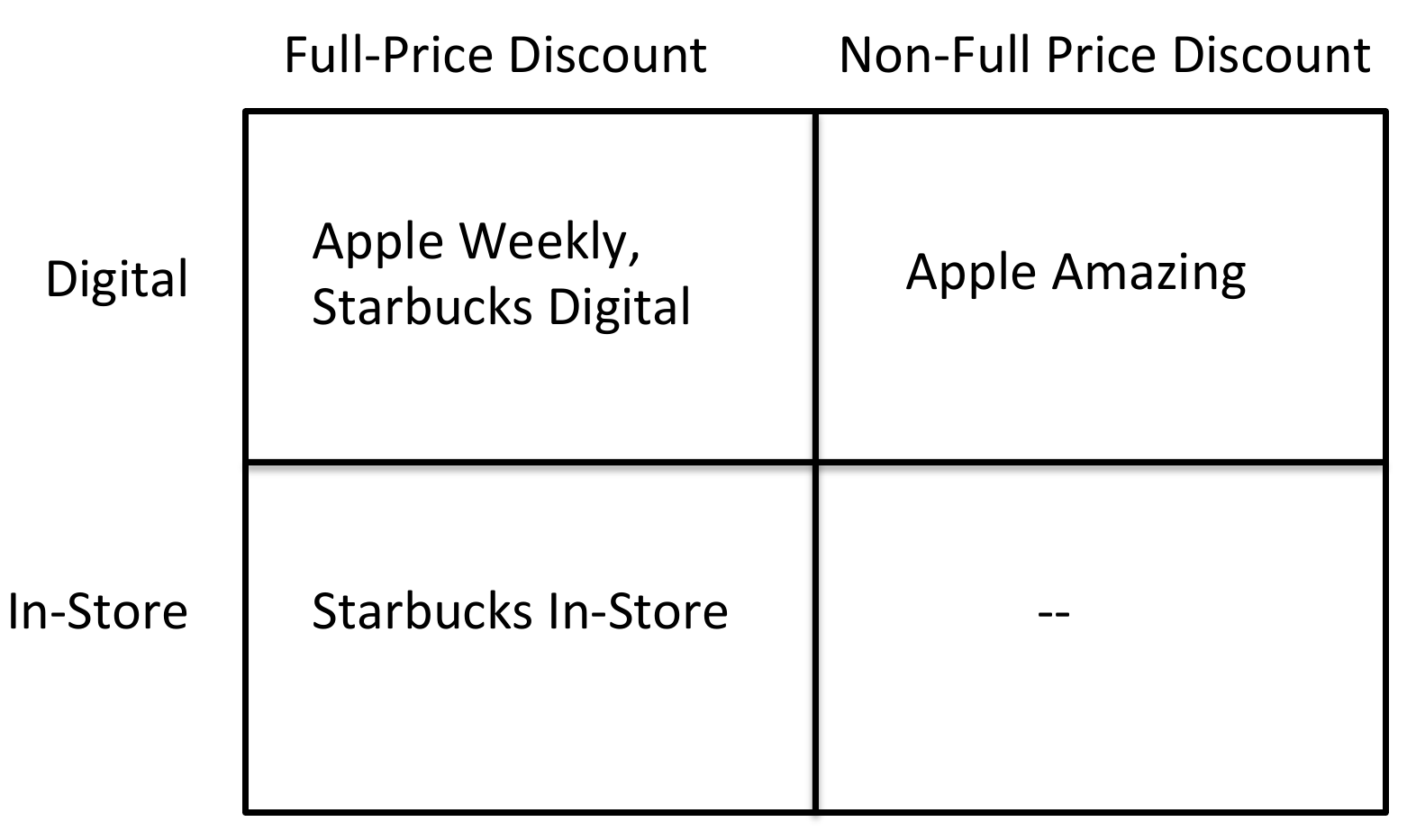}
\caption{A summary of the characteristics of the promotions we study}
\label{summary}
\end{figure}

\subsection{Related Work}
There has been extensive work on the correlation between ratings and revenue, amongst others for the case of books \citep{chevalier2006effect}, movies \citep{chintagunta2010effects,liu2006word, basuroy2003critical,reinstein2005influence,duan2008dynamics,dellarocas2005using}, games \citep{zhu2010impact} as well as new products in general \citep{cui2012effect}. \cite{luca2011reviews} and \cite{anderson2012learning} apply a regression discontinuity technique on Yelp and find that a half star (on a 5 star scale) increase on a restaurant's ratings leads to a 5-9\% higher revenue and higher sell out rates for restaurants, respectively. \cite{engstrom2014demand} apply the same technique on Google Play and find that a half star increase on ratings (on a 5 star scale) leads to 3\% more downloads for the app. Furthermore, a 10 percentile increase on the displayed number of downloads can further increase the downloads by 20\%. 
\par Another line of work tries to understand the profitability of discounts and promotions. For example, \cite{edelman2011groupon} argue that deep discounts work only if they reach customers with substantially lower valuations for the product than the regular customers. Especially relevant to our work, \cite{spriensma2012impact} estimates that being featured on Apple's App Store or Google Play can help an app gain +15 and +42 spots, respectively, on the store's top-seller rankings and \cite{carare2012impact} estimates that customers on Apple's App Store are willing to pay \$4.50 more on apps that are top-ranked than the same unranked ones. \cite{spriensma2012impact} also finds that putting apps on sale can be profitable especially if the price is cut in half or on the price points of \$0.99 or \$1.99. Hence, there are strong economic incentives for the developers to both strive to be featured on their respective stores and offer frequent discounts. \cite{ajorlou2014dynamic} even provide theoretical evidence for the optimality of a pricing strategy that drops the price to zero infinitely often in an environment where word of mouth effects are in play. Finally, \cite{adamopoulos2014social} study the effectiveness of a large scale marketing campaign carried out on Twitter and find significant returns on participating brands in terms of fan base and social media followers.
\par Our work is related to both of the aforementioned lines of research since it estimates the immediate effect of various promotions on the sales of mobile apps but also tries to understand the effect these promotions have on the ratings of the apps. The effect on ratings is important because the ratings can continue affecting the sales of the app long after the promotion is over. Our work is especially related to the work of \cite{byers2012groupon} who found that establishments that offer Groupon promotions find their Yelp rating decrease. But our work also differs from theirs in some crucial ways. First, we are studying promotions and reviews for {\it digital} and not physical goods (or services). Hence, some of the caveats that could apply for physical establishments don't apply here, e.g., all purchasers of a digital good get the exact same product no matter the price they pay, whereas owners of restaurants could discriminate against customers not paying full price. But perhaps the most important distinction between our work and the work of \cite{byers2012groupon} is that the four promotions studied in this paper are selected and offered by two third-party companies that have their own complicated incentives. Hence the selection process is not as endogenous as in platforms like Groupon. 
\par By comparing four different promotions and the different effects they had on the sales and ratings of the featured promoted apps, our work, amongst other contributions, can lead to a better understanding of how to design effective marketing campaigns.

\subsection{Data}

\par Our dataset is from the period of January 2013 to December 2015. Within this time frame, 59 and 71 apps were offered in the digital and in-store versions of the Starbucks promotion respectively, and 93 apps were offered in Apple's `App of the Week' promotion (roughly one per week). The Apple Amazing promotion is an aggregation of four promotions, that featured between them 68 unique apps.
\par Our first dataset consists of the app name, app id and the start date of the promotion for each of these 291 promoted apps. The app id is a unique string assigned by Apple to each app. In general, all apps were offered for one week each. For each promoted app, we also collect details about their characteristics such as category, price, size  and number and price of in-app purchases. We use these characteristics to understand how the promoted apps may differ from other apps in the store.
\par On the iOS App Store a user can submit two types of reviews: A star rating on a scale from 1 to 5 or a star rating and accompanied by text. All ratings contribute to a displayed average star rating but only ratings accompanied by text are displayed in the app's review page alongside with their submitted date. Between them, the 291 (59 and 71 from the Starbucks promotions and 93 and 68 from Apple's) studied promoted apps had, as of January 2016, more than 665 thousand text reviews. For each of these text reviews we collected the accompanying text, the date the review was submitted, and their star rating. We don't have data on the ratings that were submitted without accompanying text review.
\par We also use two control datasets. One consists of 56 thousand uniformly sampled apps from the set of all available apps in the AppStore. For these control apps, we collected detailed information about their characteristics such as price, size, number of in-app purchases (if any), price of in-app purchases (if any) and more. We use this control dataset to get insights about how promoted apps differ from an average app from the App Store with respect to the aforementioned characteristics. From the 56 thousand uniformly sampled apps, we further uniformly sample 5 thousand apps for which we collect all text reviews. This is our second and main control dataset which we use to control for temporal and other trends when examining the effect of the promotions on the sales and ratings of the promoted (and competing to the promoted) apps. %These 5 thousand uniformly sampled control apps have around 234 thousand text reviews. 
%Finally, we also uniformly sample 400 apps from the set of all apps that have been classified by Apple as popular for their respective category. For these two sets of 5400 apps, we collect detailed information about their text reviews (accompanying text, star rating and date of submission) as well as cumulative statistics for all their ratings. In total, the 5 thousand uniform apps have around 234 thousand text reviews and the 400 uniform popular apps have more than 1.6 million.
\par Since we can't observe the number of sales, we use the number of text reviews to make indirectly detect relative changes in the sales of an app. This is an approached used also in \cite{} for data from the Google Play store. In both Google's and Apple's store, only users that have downloaded an app can submit a review, so the number of reviews an app receives is always a lower bound on it's sales. 
\par Furthermore, we use the relative changes in the star average from text reviews to approximate relative changes in the overall star average.

\paragraph{Terminology}
Throughout the paper we will refer to any apps that were offered as part of the digital version of the Starbucks `iOS Pick of the week' promotion as \textit{Starbucks Digital apps} and to the promotion itself as \textit{Starbucks Digital}. Similarly, we will use the terms \textit{Starbucks In-Store apps} and \textit{Starbucks In-Store promotion}, \textit{Apple Weekly apps} and  \textit{Apple Weekly promotion} and \textit{Apple Amazing apps} and  \textit{Apple Amazing promotion}. There should be no confusion with apps \textit{made} by Apple or Starbucks, which are not part of any promotion discussed in this study. Finally, we will usually refer to a rating accompanied by text as \textit{text review} and simply as \textit{rating} otherwise.

\subsection{Initial Exploration}
We start by using our datasets to get insights on the general characteristics of the promoted apps and how they may differ from an average app in the App Store.
\par Table \ref{general} shows some general characteristics of the four sets of offered apps as well as of a control set of uniformly sampled apps. We see that the offered apps have a higher average price than the control and, amongst the offered apps, the ones that were offered on a non full price discount are the ones with the highest regular price. We also see that 75\% of the apps promoted in the Apple Weekly promotion offer in-app purchases, compared to 38\%-46\% for the other three promotions and only 12\% for the control. Furthermore, Apple Weekly apps offer the highest number of in-app purchases (5 compared to 2.47-2.73 for the other three promotions and 0.42 for the control) at the highest prices. This hints that the developers of the Apple Weekly apps prepare for a large influx of new users who will get the app for free and use in-app purchases to capitalize on them. In contrast, the two Starbucks and the Apple Amazing promotions have a smaller effect on the revenue of their developers since users that don't have redemption codes still pay full price for the app. Finally, the app size can provide an idea about how sophisticated the app is, for example in terms of state of the art graphics. Here too, we see that the Apple promotions feature apps that are generally much larger in size than the average app as well as the apps featured in the Starbucks promotions. This could hint toward Apple selecting and promoting state-of-the-art apps and games that, for example, take full advantage of the powerful graphics of the latest hardware releases.
\begin{table*}
\centering
%\textsc{Review Rating Distribution}\\
%\vspace{0.5cm}

\begin{tabular}{p{5cm} p{2cm} p{2cm} p{2cm} p{2.1cm}  c } 
 \toprule
 &  \textbf{Apple Weekly} &  \textbf{Starbucks In Store} &  \textbf{Starbucks Digital} & \textbf{Apple 99\cent} &\textbf{Control}\\
\midrule
Number of apps    				& 93 		& 71 		& 59 			& 68 			& 56k\\[6pt]
Mean Price 						& 2.57		& 2.78		&2.73			&3.57			& 1.21	\\[6pt]
Median Price 						& 1.99		& 2.99		&1.99			&2.99			& 0		\\[6pt]
Offers In-App Purch.				&75\% 	&43\% 	&46\% 		& 38\%		 & 12\%   \\[6pt]
{\small Avg.~\#~of~In-App~Purchases}		&  5 		& 2.47		& 2.66	 		& 2.73 		& 0.42			\\[6pt]
In-App Mean Price				& 3.85		& 0.98 	& 2.8			&2.07			& 0.62 		\\[6pt]
App Size (MB)					& 209		& 82.3		&43.7	 		&270.5		& 31.17 \\[6pt]
%Developer Size		& 24.5 (7)		& 22 (4.5)	 & 5.85 (3)		&17.1 (4)	& 50 (10)	\\[6pt]
\bottomrule
\end{tabular}
\caption{Average General Characteristics}
\label{general}
\end{table*}

\section{Effect of Promotions on Sales}\label{sec_volume}
\par One of the first metrics that can help evaluate the success of a promotion is the increase in sales it caused for the participating apps. Hence we investigate here both the immediate as well as the longer term increase that the four studied promotions caused for their featured apps.
\par Generally, we would expect low barrier promotions, i.e., free and easy redemption process, to outperform higher barrier promotions such as promotions with less trivial redemption procedures or not fully discounted. Hence, we expect the Apple Weekly and Starbucks Digital promotions to cause a larger increase than the Starbucks In-Store and Apple Weekly. Furthermore, since the Apple Weekly promotion features a banner in the front page of the App Store driving awareness, we expect the Apple Weekly promotion to outperform the Starbucks Digital one. Amongst the two high barrier promotions, we expect that the less trivial redemption procedure will outperform partially discounted promotion. 
\par
\begin{figure}[h!]
\centering
\begin{subfigure}[t]{.6\textwidth}
\centering
\includegraphics[width=\textwidth]{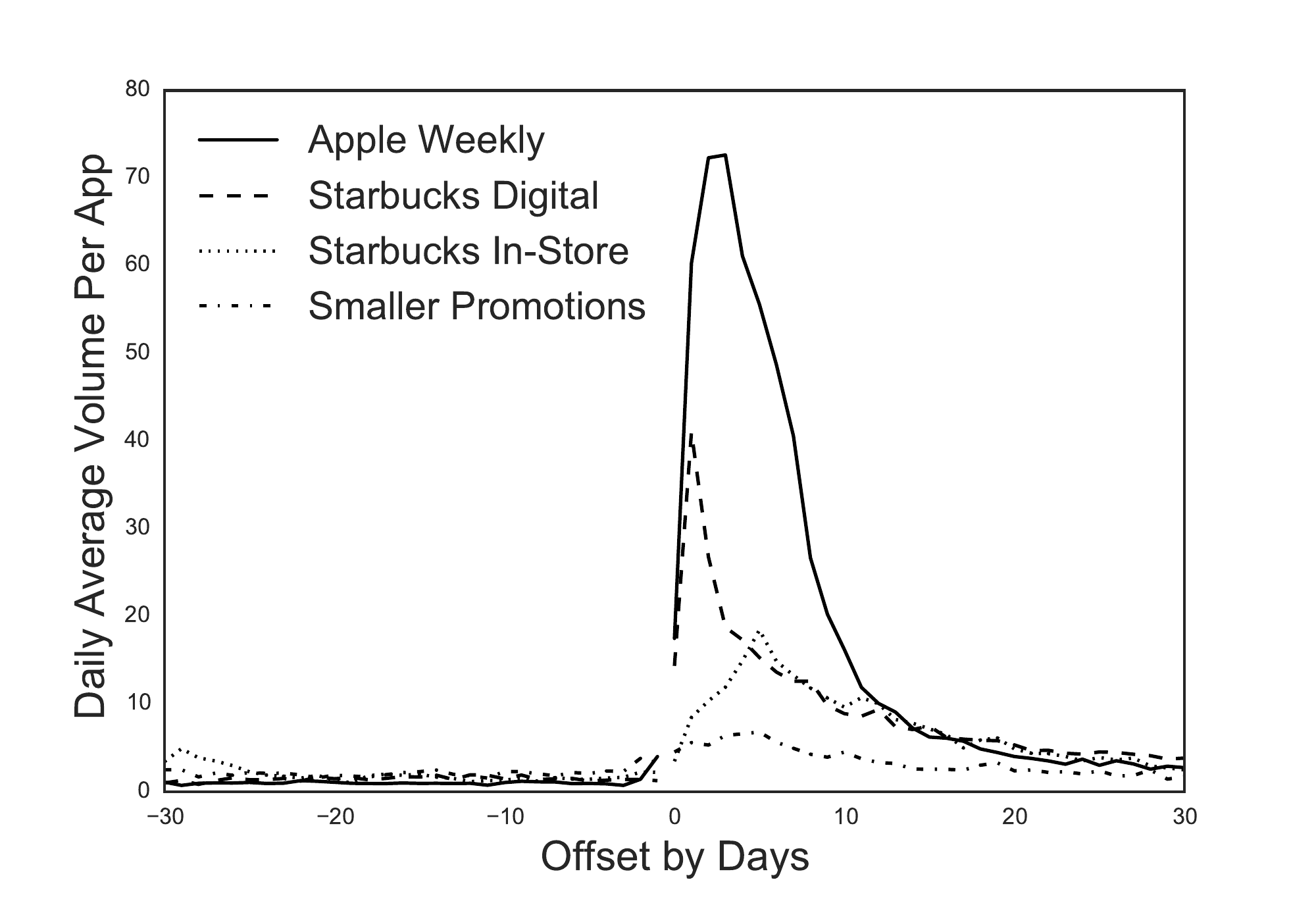}
\caption{Daily Average Review Volume, Before and After the Promotions}
\label{rev_volume}
\end{subfigure}
\quad
\begin{subfigure}[t]{.56\textwidth}
\centering
\begin{tabular}[b]{p{3.4cm} c c } 
 \toprule
				&\small{7-day Effect}	&  \small{30-day Effect} \\[6pt]
\midrule
Apple Amazing			&+52.15 (+3596\%)				& 	+18.05 (+1612\%)						\\[6pt]
Starbucks Digital			&+18.6	(+1349.3\%)				&	+8.3 (+561.6\%)							\\[6pt]
 Starbucks~In-Store 		&+10.2 (+585.7\%)				&  +5.6 (+279.4\%)						\\[6pt]
Apple Amazing		 	&+3.2 (+124.1\%)					&  +1.34 (+61.8\%)						\\[6pt]

\bottomrule
\end{tabular}
\caption{Increase in average daily volume of reviews per app. Values in parentheses represent how does the increase compare to the pre-promotion average}
\label{average_increase}
\end{subfigure}
\caption{The effect of the promotions on the sales of the apps}
\label{review_volume}
\end{figure}

 An exploratory analysis of our dataset provides support for all the predictions above. The average Apple Weekly app goes from 1.45 reviews per day, in the 7 days before the promotion starts, to 53.6 reviews per day in the 7 days of the promotion ,an increase of almost 3600\%. For the same time periods, the Starbucks Digital apps go from 1.4 to 20 reviews per day (1800\% increase), the Starbucks In-Store apps go from 1.75 to 12 reviews per day (1025\%) and the Apple Amazing apps go from 2.6 to 5.75 reviews per day (124\% increase). Table~\ref{average_increase} displays the short term (one week after and one week before the start of the promotion) and long term (one month after and one month before the start of the promotion) increase in number of daily reviews. We show both the absolute as well as percentage increase. Note that all promotions studied in this paper run for 7 days.
\par Figure~\ref{rev_volume} displays the evolution of the average daily number of reviews per app for 60 day interval offset at the beginning of the promotions, and it provides further support for our exploratory findings.

\subsection{The Econometric Model}
In order to provide statistical rigor to our descriptive findings, we estimate the following model.
\begin{equation}\label{model_volume}
\mathrm{sales}= \alpha_0 + \beta_1\mathrm{post} +\beta_2\mathrm{treat} + \beta_3\mathrm{post}\cdot\mathrm{treat}+e.
\end{equation}
The post variable is an indicator if the datapoint is after the promotion's starting date. The treat variable is an indicator if the datapoint is from an app that was part of a promotion (i.e., treatment) or not (i.e., control). The interaction variable post$\cdot$treat is the one that estimates the effect of the promotion, i.e., the amount of variation in the data that cannot be explained by the control set of reviews. The error term is $e$.
\par For the purposes of the model estimations, we restrict our data only on the period of two weeks before and two weeks after the start of the promotion. This time frame captures the effect of the promotion, while the promotion is still active as well as one week after it's over. For this time period, our dataset is comprised of 88,172 ratings for 297 treatment apps and 81,257 ratings for 2635 control apps.
\par We estimate Model \ref{model_volume} four times, once for each promotion. In doing so, we estimate effect of each promotion on the sales of the featured apps. Table \ref{regressions_volume} summarizes our estimation results. 

\begin{table*}[h!]
\begin{center}
%\textsc{}\\
\vspace{0.5cm}
\begin{tabular}{l c c c c  c} 
 \toprule
		{\it \small Promotion}	&Intercept		&post			& treat  		&  post$\cdot$ treat\\[6pt]% & R-squared \\[6pt]
\midrule
\multirow{2}{*}{Apple Weekly} 			& $5.6977^{***}$			&$-0.7120$		&$-2.3412$	& $35.6709^{***}$	\\[6pt]	%		&0.04			\\[6pt]
											&(0.924)				&(1.295)		&(2.208)	&(2.671)			\\[6pt]

\midrule

\multirow{2}{*}{Starbucks Digital} 	&  $4.8775^{***}$	& $0.1978$			&$0.5511$ 		& $12.6284^{***}$\\[6pt]	%		&0.13			\\[6pt]
											&(0.613)			&(0.852)			&(1.589)		&(1.897)\\[6pt]

\midrule

\multirow{2}{*}{Starbucks In-Store} 	& $5.6487$		& 	$1.7257$			&$-1.5137$ 		& $8.2381^{**}$	\\[6pt]%		&0.04		\\[6pt]
									&(0.849			&(1.170)				&(2.053)			&(2.540)		&\\[6pt]
\midrule

\multirow{2}{*}{Apple Amazing} 				&  $5.7345^{***}$	& $-0.1381$		&$-0.5999$ 		& $2.5726^{**}$	\\[6pt]%	&0.003			\\[6pt]
									&(0.295)			&(0.404)			&(0.620)			&(0.813) 	\\[6pt]%	&\\[6pt]

%\midrule
%\multirow{2}{*}{Number of Tags} 	& $0.4500^{***}$	& $ -0.0526^{***}$		&$0.0008$ 		& $0.0002$			&0			\\[6pt]
	%								&(0.008)			&(0.05)					&(0.001)		&(0.001)		&\\[6pt]

\bottomrule
\end{tabular}
\end{center}
Values in parentheses are standard errors.\\
\small{$^{*}:p<0.05$, $^{**}: p<0.01$, $^{***}:p<0.001$}
\caption{Effect of promotions on sales}
\label{regressions_volume}
\end{table*}

\subsection{Results}
\par We discuss here the results from the estimation of Model \ref{model_volume} and how they relate to our descriptive results.
\par The fourth column of Table \ref{regressions_volume} shows the coefficients for the post$\cdot$treat variable, which estimates the effect of the promotions, and it confirms our descriptive findings, shown in Figure \ref{rev_volume}. 
\paragraph{Apple Weekly} As we expected, Apple Weekly has the largest coefficient for the post$\cdot$treat variable confirming that it's the promotion with the largest positive impact on the featured apps' sales. This can be explained not only by the low barrier nature of the promotion but also by the large awareness that the banner in the front page of the App Store brings. In addition to these factors, the app is promoted and appears to be endorsed by Apple itself, hence providing users with further confidence. As seen from Figure \ref{rev_volume} the large effects last around 2 weeks, after which a gradual descent `back to normal' begins. The promotion is only one week in duration, and hence we believe that the extra days of increased volume of sales is due to a combination of word-of-mouth as well as higher ranks in the top selling charts, which came as a result of the promotion.
\paragraph{Starbucks promotions} As can be seen in the second and third row of Table \ref{regressions_volume}, the coefficients of post$\cdot$treat for both of the Starbucks promotions are positive and highly significant, hence confirming the positive increase in sales we observed in Figure \ref{rev_volume}. Moreover, the coefficient is larger for the Starbucks Digital version compared to the Starbucks In-Store, confirming that low barrier promotions attract larger increases than higher barrier ones. Figure \ref{rev_volume} also shows that the increase in the digital promotion is much more sudden than the in the in-store promotion. This can be because customers redeeming a coupon for the in-store promotion can pick up the coupon from the physical locations but can wait up to three months before going through the process. In contrast, customers can redeem the digital promotion, right from their phones, only within the week of the promotion.
\par Similar to the Apple Weekly promotion, we expect that the Starbucks promotions caused an increase in word-of-mouth for the featured apps and helped them climb up the ranks of the top charts, which can explain why there are still some abnormally large sales (compared to before the promotion) even after the 7 days of the promotion were over. 

\paragraph{Apple Amazing} Even though the Apple Amazing promotion is the only one not offering a full price discount, as can be seen in the fourth row of Table \ref{regressions_volume}, it still observes a positive and significant increase in sales. Confirming what we observed in Figure \ref{rev_volume}, this increase is the smallest amongst the studied promotions. Smaller increase in sales means smaller climb in the top-charts ranks which can also explain why this promotion seems to be the one returning to their normal pre-promotion sales faster from the other ones.

\par The results from this section show that low barrier promotions, i.e., digital and free, have the largest and most immediate increase in their sales whereas higher barrier promotions, i.e., partial discount or offline redemption procedure, have smaller and less sudden, but still substantial, increases.

\section{Effect of Promotions on Ratings}
\par Having studied the effect of the promotion on sales, we turn our attention in the effect on ratings. A promotion is usually designed to attract users that would otherwise not buy a product with the intention to turn them into long-term paying customers. Does this increase in sales need always come with the risk of lower ratings, as observed in \cite{byers2012groupon}, or can practitioners mitigate undesired effects by designing the promotion carefully? An extended literature has shown the importance of positive ratings in the economics success of products and services \citep{}, hence any effects that the promotions will have on the ratings of featured products and services, can continue affecting their success long after the promotions are over. 
\par Even though our work is very similar in spirit with the work of \cite{byers2012groupon}, our setting has important differences. Perhaps the most important is that the promoted apps are selected by Apple or Starbucks, two companies with their own complicated objectives and incentives, that may not always align entirely with the objectives and incentives of the app developers. Unlike settings such as Groupon, a developer cannot just add their app in the Apple Weekly promotion, it needs to be selected (or at least agreed by) Apple. This induces a selection procedure that makes our predictions for the effects on the ratings harder. Take for example the Apple Weekly promotion. As can be seen in Figure~\ref{rev_volume}, Apple Weekly is the promotion that causes the largest increase in sales. This large influx of new users can be risky since some of them may not be in the target group of the promoted app. On the other hand, the promotion is on the front page of the AppStore and that means that the users that are exposed to the banner are customers that are actively browsing for apps to download. The fact that Apple is endorsing the promoted apps can also induce some positive `social influence'-type bias.
\par An exploratory analysis of the immediate effects of the promotions on the ratings of the promoted apps, shows that the apps featured in the Apple Weekly promotion experience, on average, an increase of 0.4 stars, from 3.86 in the week before the promotion starts to 4.27 for the week after the promotion started. In the same time period, apps featured in the Starbucks Digital promotion doesn't seem to have been affected, with their ratings increasing only by 0.01 star (from 4.21 to 4.22). The ratings for apps featured in the Starbucks In-Store promotion increased by 0.14 star (from 4.33 to 4.47) and for apps featured in the Apple Amazing promotion decreased by 0.34 star (from 4.3 to 3.96).
\par
\begin{figure*}
\begin{center}
\begin{subfigure}[t]{\textwidth}
\centering
\begin{subfigure}[t]{0.35\textwidth}
\centering
\includegraphics[width=\textwidth]{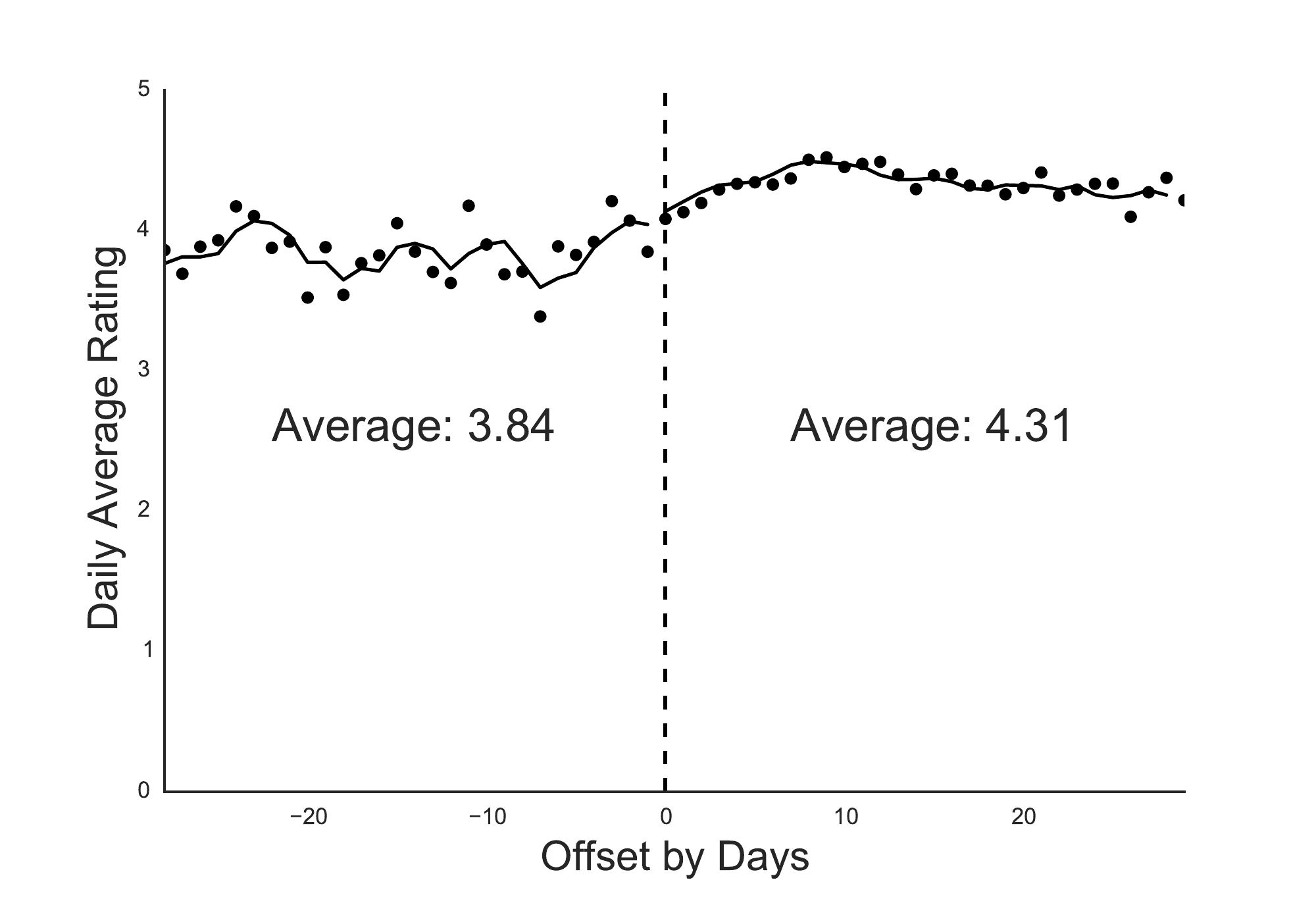}
\end{subfigure}
\quad
\begin{subfigure}[t]{0.35\textwidth}
\includegraphics[width=\textwidth]{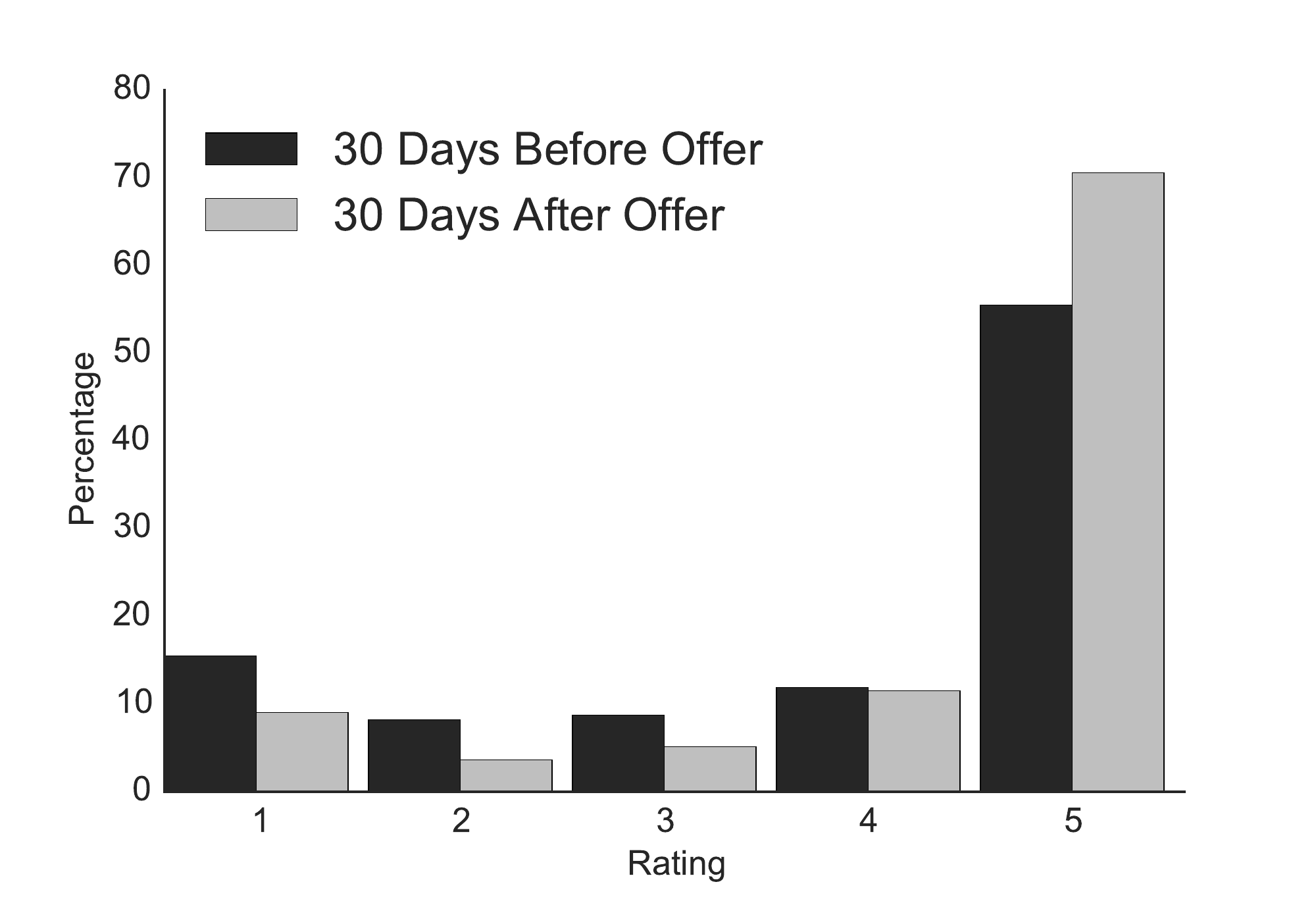}
%\caption{\tiny{Apple Weekly}}
%\label{apple_dist_bfr_aft}
\end{subfigure}
\caption{Apple Weekly}
\label{apple_rating_evol}
\end{subfigure}
\quad
\begin{subfigure}[t]{\textwidth}
\centering
\begin{subfigure}[t]{0.35\textwidth}
\centering
\includegraphics[width=\textwidth]{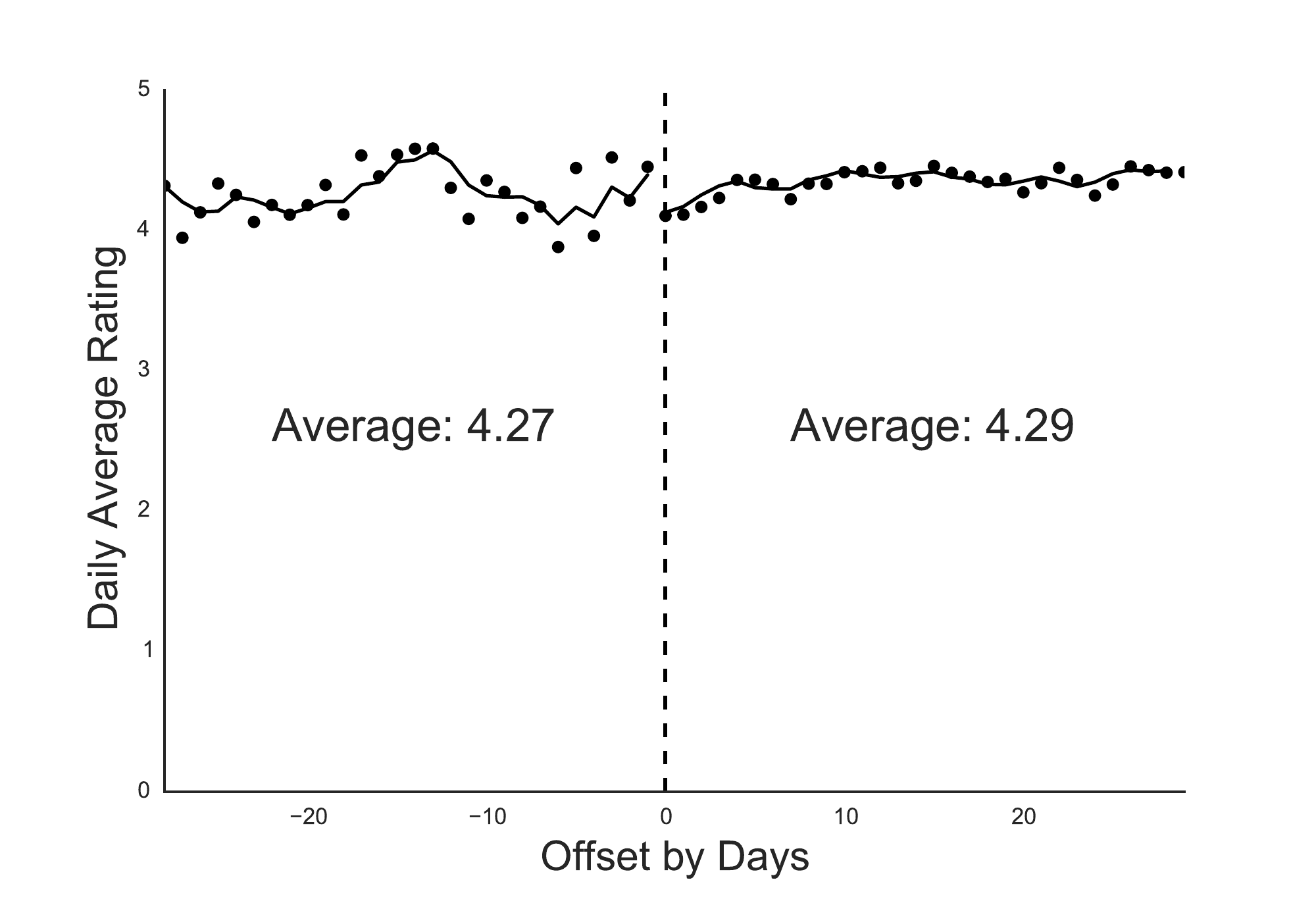}
\end{subfigure}
\quad
\begin{subfigure}[t]{0.35\textwidth}
\includegraphics[width=\textwidth]{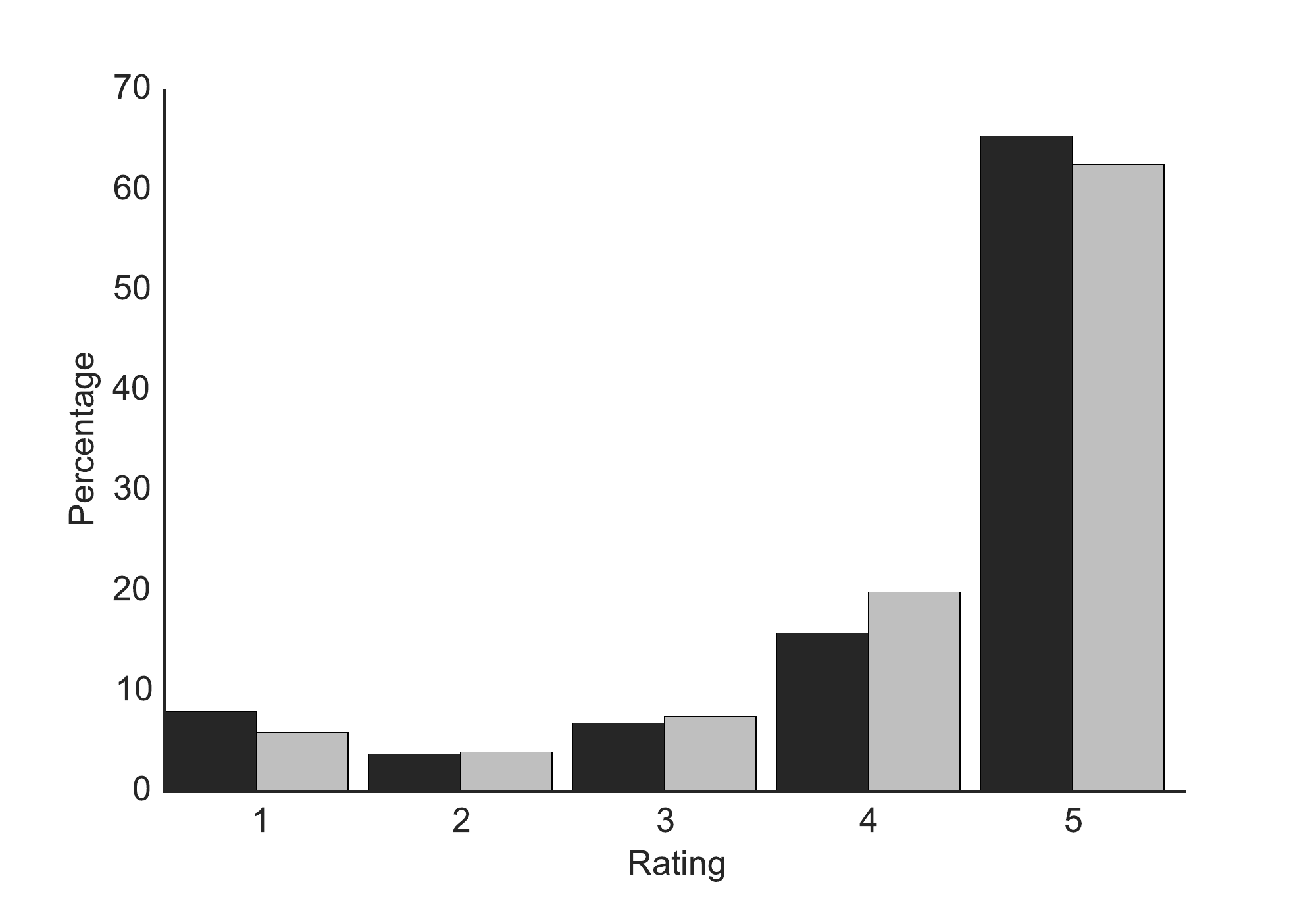}
\end{subfigure}
\caption{Starbucks Digital}
\label{starb_dig}
\end{subfigure}
\quad
\begin{subfigure}[t]{\textwidth}
\centering
\begin{subfigure}[t]{0.35\textwidth}
\centering
\includegraphics[width=\textwidth]{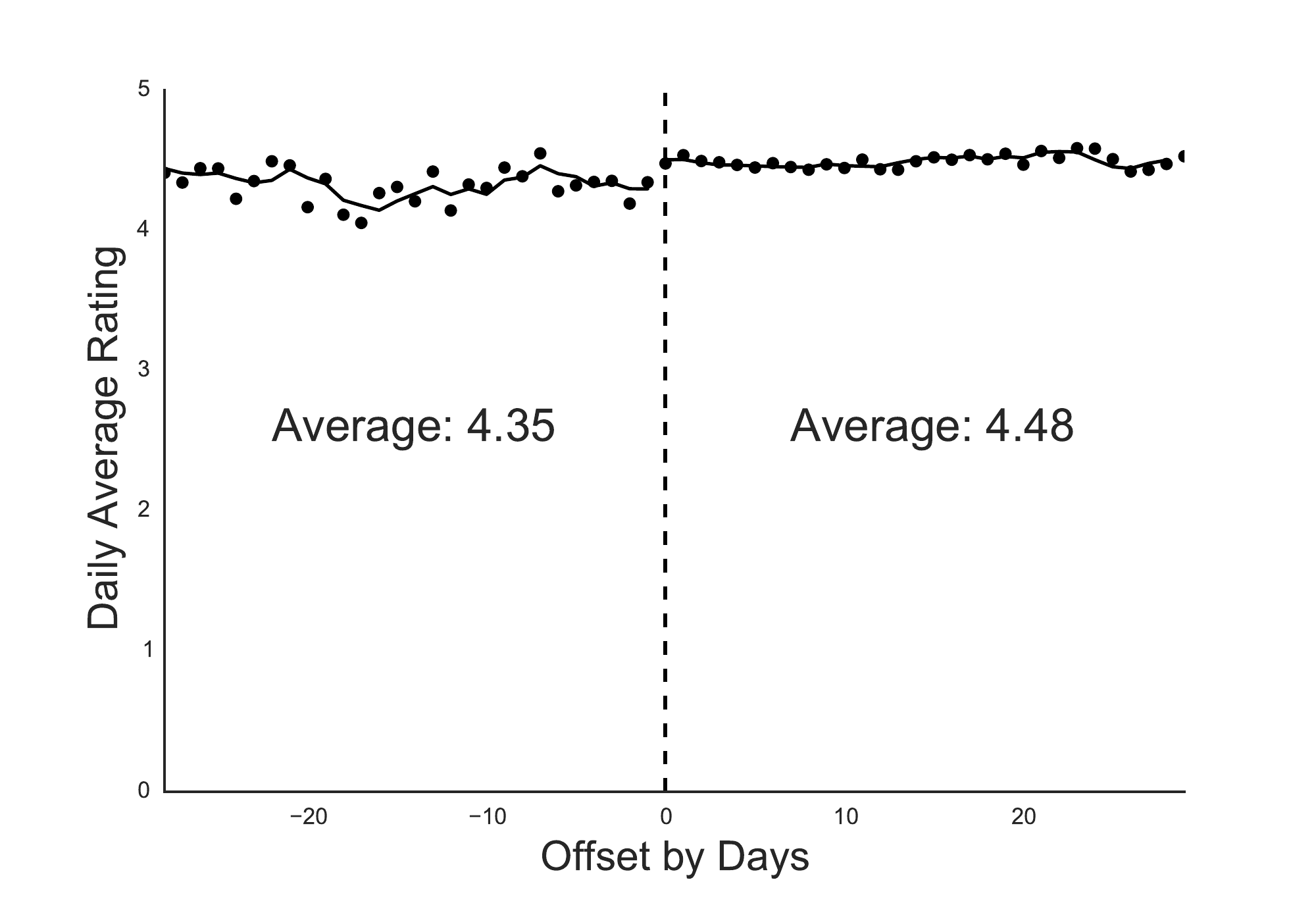}
%\caption{Evolution of Ratings for Starbucks - Digital Apps}
\end{subfigure}
%\caption{Evolution of Ratings for Starbucks - In StoreApps}
%\label{apple_raring_evol_200}
%\end{subfigure}
\quad
\begin{subfigure}[t]{0.35\textwidth}
\includegraphics[width=\textwidth]{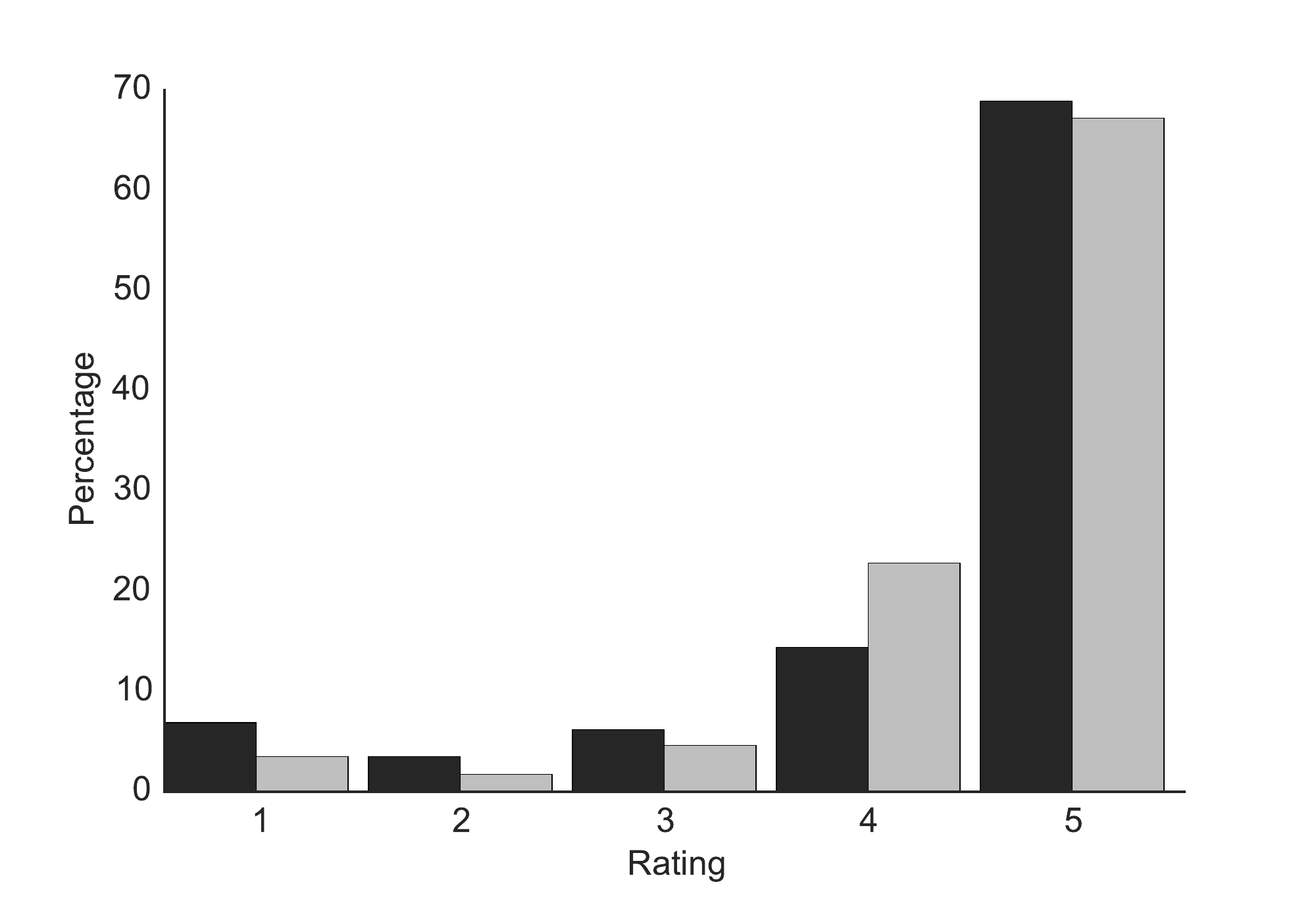}
%\caption{\tiny{Starbucks Digital Download}}
%\label{starb_dist_bfr_aft}
\end{subfigure}
\caption{Starbucks In-Store}
\label{starb_in_store}
\end{subfigure}
\quad
\begin{subfigure}[t]{\textwidth}
\centering
\begin{subfigure}[t]{0.35\textwidth}
\centering
\includegraphics[width=\textwidth]{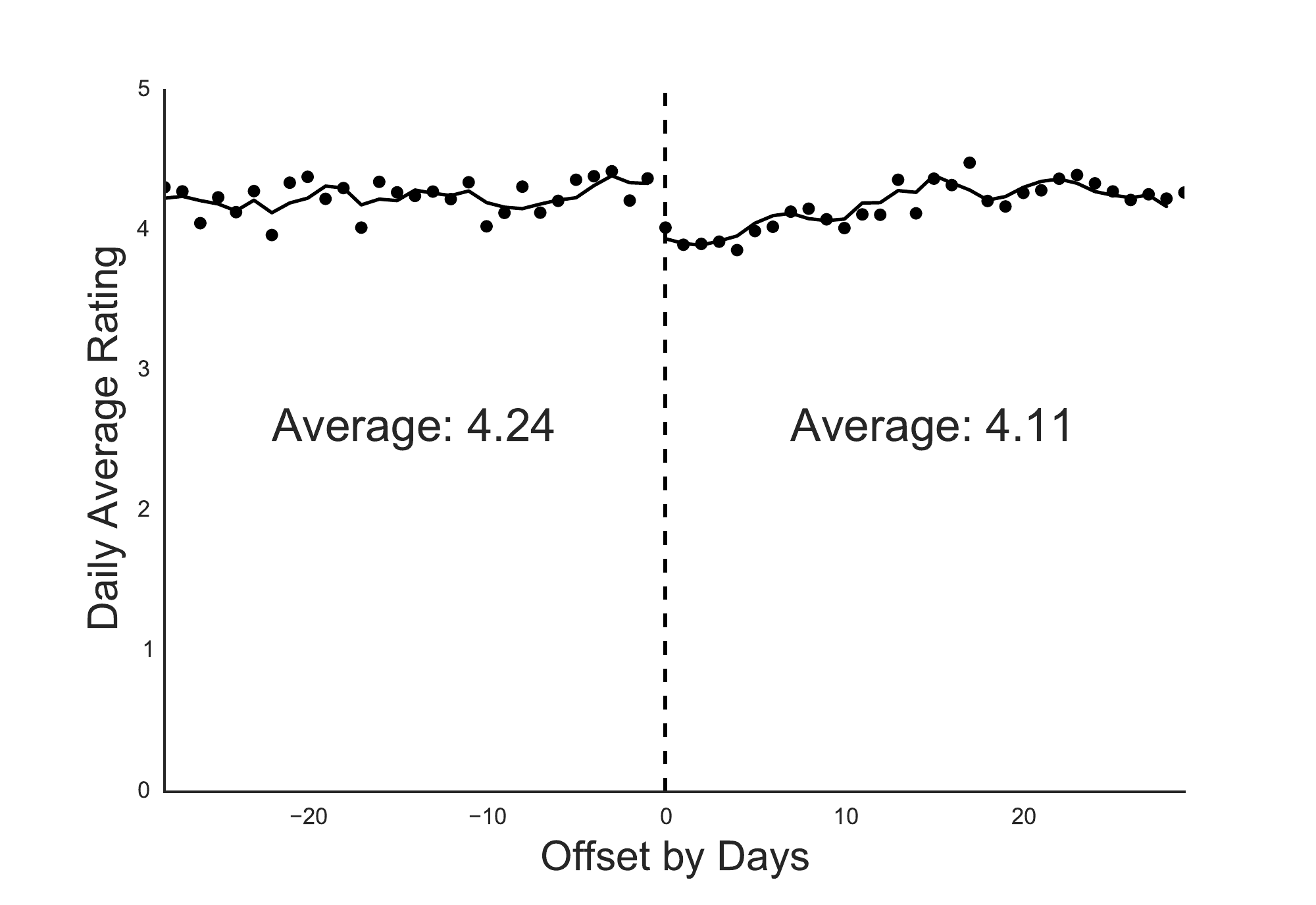}
\end{subfigure}
\quad
\begin{subfigure}[t]{0.35\textwidth}
\centering
\includegraphics[width=\textwidth]{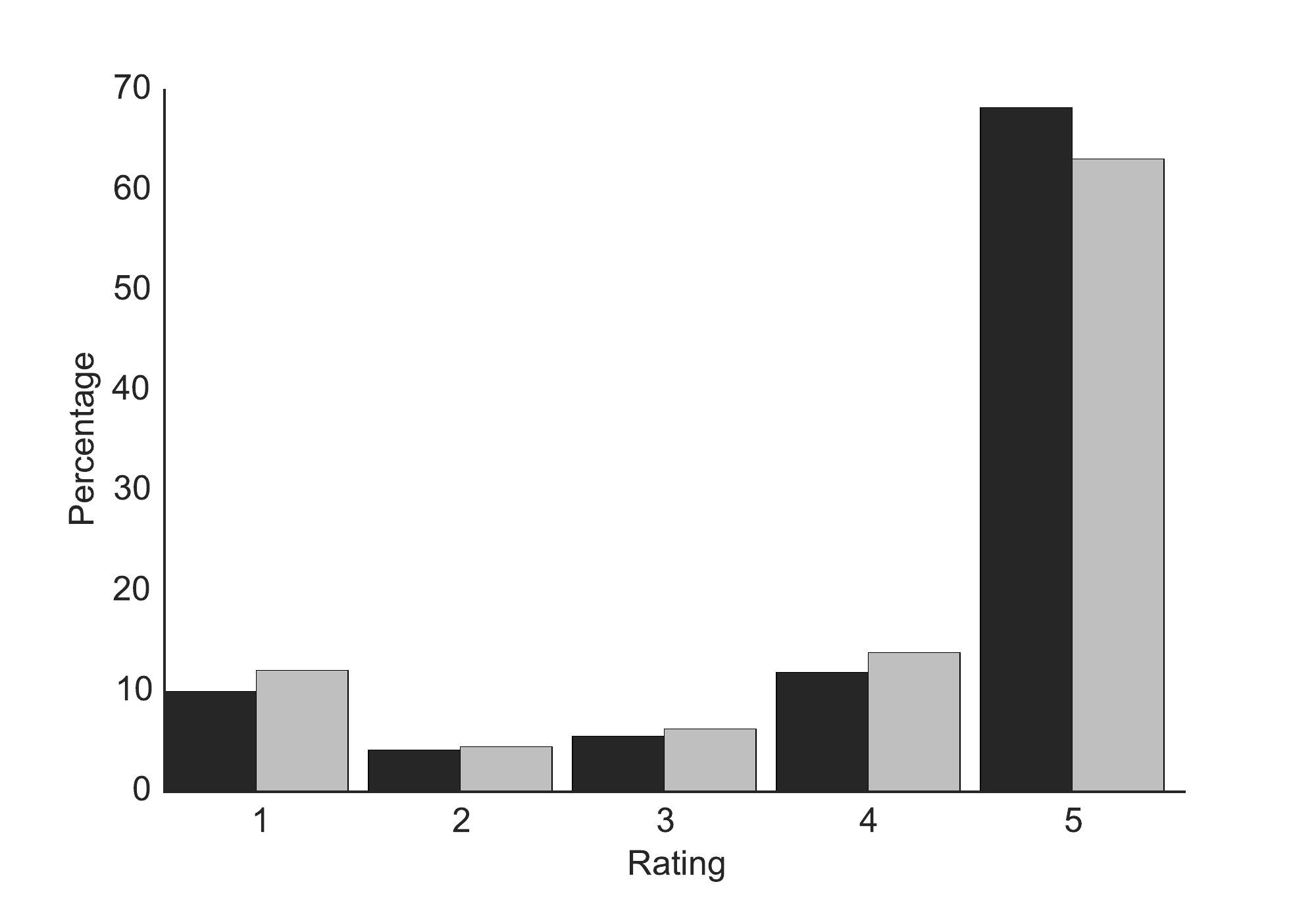}
\end{subfigure}
\caption{Apple Amazing}
\label{apple_am}
\end{subfigure}
\end{center}
\textit{On the Left:} Evolution of daily average rating for a two month period centered around the offer date (dotted vertical line) for the four studied promotions.\\ \textit{On the right:} the distribution of ratings for the same time period.
\caption{Daily Average Star Rating, Before and After the Promotions}
\label{per_rank_evolution}
\end{figure*}

 Figure \ref{per_rank_evolution} shows a more detailed view on the effect that the promotions had on the ratings. On the left hand side, is the evolution of the daily average rating for the two month period offset at the start of the promotion (all promotions run for 7 days) alongside the average rating for each 30 day period. On the right hand side, is the actual distributions of the ratings for each 30 day period (before and after the start of the promotion). Figure \ref{per_rank_evolution} provides further support for all the descriptive statistics we mentioned above: ratings become substantially more positive for the Apple Weekly apps, remain stable or go slightly up for the Starbucks promotions and there is a sudden decrease for the Apple Amazing apps, but one that seems to be partially corrected after the end of the promotion.

\subsection{The Econometric Model}
We estimate the following model in order to provide statistical support for the descriptive results shown in Figure \ref{per_rank_evolution}.
\begin{equation}\label{model_2}
\mathrm{rating}= \alpha_0 + \beta_1\mathrm{post} +\beta_2\mathrm{treat} + \beta_3\mathrm{post}\cdot\mathrm{treat}+e
\end{equation}
The post variable is a binary indicator if the datapoint is before the start date of the promotion and the treat variable is a binary indicator if the datapoint is from an app that was part of a promotion. The interaction variable post$\cdot$treat is the effect that is unique to the promoted apps after the promotion started. Hence, the coefficient of that variable estimates the effect on the promoted apps that cannot be explained by the control dataset, i.e., the causal effect of the promotion. The error term is $e$.
\par As we did with the regressions for the effect on sales, we restrict our attention on a period of two weeks before and two weeks after the start date of the promotions. This makes the time window long enough to capture the immediate effects of the promotions while the promotions are in progress as well as one week after they are over. For this time frame, our dataset is comprised of 88,172 ratings for the 297 treatment apps and 81,257 ratings for 2635 control apps.
\par We estimate Model \ref{model_volume} four times, once for each promotion and summarize our results in Table \ref{regressions_1}

\begin{table*}[h!]
\begin{center}
%\textsc{}\\
\vspace{0.5cm}
\begin{tabular}{l c c c c  c} 
 \toprule
		{\it \small Promotion}			&Intercept 			&post 					&treat 			&  post$\cdot$treat\\[6pt]% & R-squared \\[6pt]
\midrule
\multirow{2}{*}{Apple Weekly} 					& $4.0104^{***}$	&$-0.01084$			&$-0.1458^{**}$	& $0.4167^{***}$	\\[6pt]	%		&0.04			\\[6pt]
												&(0.017)			&(0.024)				&(0.046)			&(0.050)			\\[6pt]
	
\midrule

\multirow{2}{*}{\small Starbucks Digital} 	&   $4.1758^{***}$	& $-0.0029$	&$0.052$ 		& -0.001\\[6pt]	%		&0.13			\\[6pt]
											&(0.021)			&(0.03)				&(0.05)		&(0.05)		&\\[6pt]

\midrule

\multirow{2}{*}{Starbucks In-Store} 		& $4.4737^{***}$		& 	$0.0307$			&$-0.1369^{***}$ 		& $0.11^{***}$	\\[6pt]%		&0.04		\\[6pt]
										&(0.014)					&(0.018)			&(0.037)				&(0.04)		&\\[6pt]
\midrule

\multirow{2}{*}{Apple Amazing} 					&  $4.2258^{***}$			& $-0.1043^{***}$	&$0.0769$ 		& $-0.2320^{***}$	\\[6pt]%	&0.003			\\[6pt]
										&(0.022)					&(0.029)			&(0.045)			&(0.054) &	\\[6pt]%	&\\[6pt]

%\midrule
%\multirow{2}{*}{Number of Tags} 	& $0.4500^{***}$	& $ -0.0526^{***}$		&$0.0008$ 		& $0.0002$			&0			\\[6pt]
	%								&(0.008)			&(0.05)					&(0.001)		&(0.001)		&\\[6pt]

\bottomrule
\end{tabular}
\end{center}
Values in parentheses are standard errors.\\
\small{$^{*}:p<0.05$, $^{**}: p<0.01$, $^{***}:p<0.001$}
\caption{Effect of the promotion on the short term ratings of the featured apps.}
\label{regressions_1}
\end{table*}

\subsection{Results}
We discuss here the results of our statistical analysis and how they tie up with our descriptive results.
\par The coefficient for post$\cdot$treat when Model \ref{model_2} is run on the Apple Weekly dataset is positive and highly significant. This confirms that our descriptive findings, that the Apple Weekly promotion has a positive effect on the ratings of the featured apps, is statistically significant. Figure \ref{apple_rating_evol} also shows that the positive effect in the ratings seems to be long term, one that remains even after the promotion is over. A look at the distributions of the ratings shows a substantial increase of 5-star ratings and decrease of 1, 2 and 3-star ratings, further confirming the shift towards more positive ratings. 
\par The estimations of the coefficient for the post$\cdot$treat variable for the Starbucks promotions, shown in the second and third row of Table \ref{regressions_1}, provide statistical confirmation for our descriptive results about the effect of these promotions had on the ratings of the featured apps. For the digital version of the promotion the coefficient is not significantly different from zero but for the in-store version, the coefficient is positive and significant. 
A look at the evolution of the daily average rating for the Starbucks Digital apps, in Figure \ref{starb_dig}, shows that that ratings didn't shift very much but seem to have been stabilized, perhaps because of the increased volume of reviews. The changes in the distributions of the ratings show a slight decline in 1 and 5-star ratings and a slight increase in 4-star ratings. For Starbucks In-Store, Figure \ref{starb_in_store} ,the average ratings seem to increase slightly and stabilize. Again, this stabilization could be because of the increased volume of reviews. Moreover, a comparison of the distributions of the ratings before and after the start of the promotion, show an increase of 4-star and decrease of 1-star ratings.
\par The in-store promotion involves a customer noticing the printed coupon in a physical location, picking it up and then redeeming it on their phone. In contrast, the digital version of the promotion offers the users the ability to claim their free app with only two taps from the within the Starbucks iOS app. Hence, we ask, what types of customers are more likely to go through each process? Customers are more likely to go through with the offline redemption procedure if they are positively predisposed towards the promoted app, e.g., they are aware of the app from their friends or from some other source. Hence, this small barrier seems to be effective in filtering out users that wouldn't be in the target audience of the group and wouldn't enjoy it. 
\par Even though the Starbucks Digital and Apple Weekly promotion are both digital and free, only Apple's promotion causes an increase in the ratings. This could be due to few factors. First, it could be that the selection procedure of Apple is such that it discovers high quality apps and presents them to their users. From a private conversation the author had with a person in charge of the Starbucks promotions, the selections process for the Starbucks apps seems to be essentially the same as the one for the Apple promotions, with Apple needing to confirm the selections made by Starbucks before they go live on either of the two Starbucks promotions. Hence, another factor is the incentives and goals of each company. Starbucks seems to be offering the promotion as an additional service to their customers and our data show that they tend to choose apps that are higher rated than the Apple Weekly apps. This could perhaps be because Starbucks wants to offer their customers already popular and risk-free apps, whereas Apple could have incentives to choose more niche apps, in order to help their customers discover even more great apps than what they already know. Finally, users that are exposed to the Apple Weekly promotion are users that are already browsing the AppStore hence they are already actively looking for apps to purchase. This seamless and well timed integration of the promotion in the user experience could be why we see the different effects in the ratings of their featured apps. Moreover, the apps featured in the Apple Weekly promotion are promoted and endorsed by Apple itself, hence an iOS user could be predisposed positively towards them.
\par The coefficient of the post$\cdot$treat variable for the Apple Amazing promotion, is highly significant and the only, out of the four studied promotions, that is negative. This is in agreement with our descriptive results, shown in Figure \ref{apple_am}, that show a sudden decrease in the ratings immediately following the start of the Apple Amazing promotion. Figure \ref{apple_am} also shows that this decrease seems to be partially corrected as time goes on after the end of the promotion. A look at the distributions of the ratings before and after the promotion shows a decrease in 5-star and increase in 1-star ratings hence further confirming the shift towards lower ratings.
\par This decrease must have something to do with the fact that the Apple Amazing promotion is the only one not offering a full discount. Even though the discount is around 75\%, with the average normal price for the Apple Amazing promotions being around \$4, the \$0.99 price tag is still higher than the median (\$0) and only slightly lower than the mean (\$1.2) app price in the App Store. Note that users that are exposed to the Apple Amazing promotion don't get information about the app's original price. Hence, from a customer's point of view they are offered an app that in comparison to the average app in the store, is only slightly cheaper. Hence, the promotion and the endorsement by Apple may have incentivized some users to purchase the app even though they would not have buy otherwise. The price-tag may have then caused these users to be dissatisfied and submitting a low rating.
\par Note that the Apple Amazing promotion is the one most similar to promotions on platforms like Groupon, because it's the only one where the apps are not given out for free. It's interesting that it's the only promotion we find to have a negative effect on the ratings. We think this is a finding that can contribute in further understanding of the `Groupon phenomeon' \citep{byers2012groupon}, and one that practitioners should take into account when designing their promotions.

%\subsection{Differences in Differences}
%\subsubsection{Review Volume}
%\begin{figure}
%\centering%
%\begin{tabular}[b]{| p{4cm} | c | r | r | r | }
%\hline
%& \tiny{Overall D-in-D  Short} & \tiny{Overall D-in-D  Long} & \tiny{Similar D-in-D Short} & \tiny{Similar D-in-D Long} \\\hline
%Apple Pick &  5824  &579 & 7284 & 615  \\\hline
%Starbucks in Store & 1414   & 221 & 1643 & 370 \\\hline
%Starbucks Digital  & 812  & 111 & 872 &195\\\hline
%Apple Amazing &154 & 693  &  35& 436\\\hline
%\end{tabular}
%\caption{Effect on Review Volume}
%\label{dnd_volume}
%\end{figure}

%\subsubsection{Rating}
%\begin{figure}
%\centering%
%\begin{tabular}[b]{| p{4cm} | c | r | r | r | }
%\hline
%& \tiny{Overall D-in-D  Short} & \tiny{Overall D-in-D  Long} & \tiny{Similar D-in-D Short} & \tiny{Similar D-in-D Long} \\\hline
%Apple Pick &  0.58  & 0.73 & 1.14 & 0.69  \\\hline
%Starbucks in Store &  -0.39  & -0.05 & -0.43 & -0.03  \\\hline
%Starbucks Digital  & -0.49 & 1.3 & 0.23 & -0.47\\\hline
%Apple Amazing &0.01 & -0.12  & 0.3 & 0.19\\\hline
%\end{tabular}
%\caption{Effect on Rating}
%\label{dnd_rating}
%\end{figure}

\section{Effect of App Characteristics}
\par All the studied promotions caused an increase in the sales of the promoted apps, but was the extend of this increase correlated with any of the characteristics of the app\footnote{The author would like to thank the anonymous CIST reviewer that suggested this research idea}? For example, when offered for free, do apps that are normally more expensive benefit more or less from apps that are cheaper? How about the previous popularity of the promoted apps? Do apps that are already popular and established benefit more or less from a promotion? Insights towards this direction can help practitioners understand what type of apps can benefit the most from each type of promotions, and design their marketing strategy accordingly.
\par In order to examine this question, we define a new variable, called {\it boost}, as the difference between the average number of reviews an app received during the 7 days of the promotion and the 7 days before the promotion started.
\par We focus on four characteristics of an app: popularity before the start of the promotion, normal price, age (in days),  and size (in Mega Bytes). Popularity before the start of the promotion will help us see if there are any `rich-get-richer' effects, where apps that were already popular benefit disproportionally from price reductions and promotions. The regular price of the app can capture if, amongst apps that are offered for the same price (free or otherwise), the apps that are more expensive normally will attract larger interest. For example, more users that held back their purchases for more expensive apps can now have the opportunity to get it at a discount. Furthermore, the regular price can also be correlated with the quality of the app. The age of the app can capture if an app is established and had more time to build their user base and word of mouth, and the size in MegaBytes of the app can signal the level of sophistication that app has, e.g., state of the art graphics or many levels for a game.
\subsection{Econometric Model}
\par Since we don't always have the date that the first version of an app appeared in the AppStore, we approximate the age of an app by calculating the number of days between the start of the promotion the app was featured in, and the date of the {\it first review} it received. Then we calculate the number of reviews that an app received during its lifetime up to the start of the promotion. We then divide this number by the app's age to get the average daily number of reviews for that app up to the start of the promotion. We use this metric as an indicator for the popularity of the app before the start of the promotion. The size and price are as observed at the time of scraping, which was January 2016. Even though, there may be some differences between the observed size and price and that of the offered version, we believe that they still provide a reasonable approximation to the characteristics of the offered version.

\par In order to find correlations between the boost and the above characteristics, we estimate the following model

\begin{equation}\label{model_4}
\mathrm{boost}= \alpha_0 + \beta_1\mathrm{POP} + \beta_2\mathrm{age}+\beta_3\mathrm{size} +\beta_4\mathrm{price}+e,
\end{equation}
where POP is the app's previous popularity, as calculated above, and age, size and price the characteristics of the app as discussed above. The error term is $e$. We estimate Model \ref{model_4} four times, once for each promotion and our results are summarized in Table \ref{regressions_characteristics}. 
\begin{table*}[h!]
\begin{center}
%\textsc{}\\
\vspace{0.5cm}
\begin{tabular}{l c c c c  c} 
 \toprule
{\it \small Promotion}				&Intercept				&POP				& age  			& size	 & price\\[6pt]	% & R-squared \\[6pt]
\midrule
\multirow{2}{*}{Apple Weekly} 	& $557.7493^{***}$		&$10.5821$			&$-0.2063$		&0.0765		 &-30.8789\\[6pt]	%		&0.04			\\[6pt]
								&(135.980)				&(15.860)			&(0.174)		&(0.177	)	&(32.711)	\\[6pt]

\midrule

\multirow{2}{*}{Starbucks Digital} 		& $134.03$		& 	$11.18$		&$0.01^{***}$ 		& -0.08 	&-6.00	\\[6pt]%		&0.04		\\[6pt]
									&(97.27)			&(6.18)			&(0.09)			&(1.4)	& (15.01)	\\[6pt]

\midrule

\multirow{2}{*}{\small Starbucks In-Store} 	&   $84.3671$	& $14.2763^{**}$	&$-0.0105$ 		&-0.04 	&-11.99\\[6pt]	%		&0.13			\\[6pt]
										&(58.85)		&(4.33)			&(0.07)		&(0.30)	&(17.17)\\[6pt]

\midrule
\multirow{2}{*}{Apple Amazing} 						& $-1.13$		& $7.30^{***}$		&$1.29^{*}$  	& $0.002$ &5.43	\\[6pt]%	&0.003			\\[6pt]
										&(14.93)		&(1.29)			&(0.01)		&(0.01) 	&(3.36)	\\[6pt]%	&\\[6pt]

%\midrule
%\multirow{2}{*}{Number of Tags} 	& $0.4500^{***}$	& $ -0.0526^{***}$		&$0.0008$ 		& $0.0002$			&0			\\[6pt]
	%								&(0.008)			&(0.05)					&(0.001)		&(0.001)		&\\[6pt]

\bottomrule
\end{tabular}
\end{center}
Values in parentheses are standard errors.\\
\small{$^{*}:p<0.05$, $^{**}: p<0.01$, $^{***}:p<0.001$}
\caption{Effect of app characteristics in the success of promotions}
\label{regressions_characteristics}
\end{table*}

\subsection{Results}
We discuss here the results of the estimations of Model \ref{model_4}.

\paragraph{Regular Price} Unlike promotions on platforms like Groupon, none of the promotions we study in this paper display the normal price of the featured app\footnote{We noticed only one case where Apple displayed the normal price, of \$9.99, for an app featured in the promotion.}. Hence we expect users to not be sensitive to that information but, instead, only on the price they are called to pay. Indeed we see in Table \ref{regressions_characteristics} that the coefficient of the variable price is not significantly different from zero for any of the four studied promotions. This means that no matter the regular price, the featured apps in the studied promotions benefit, on average, the same as the other apps featured in the same promotion.
\par Practitioners should be aware that even if they are cutting down their margins more than other apps featured in the same promotion, as long as the final discounted price is the same, they will not receive any advantage during the promotion.

\paragraph{Previous Popularity} Table \ref{regressions_characteristics} shows that the previous popularity of an app doesn't matter when the promotion is low barrier (i.e., digital and free) but it does when it's high barrier (i.e., offline redemption or not free). In low barrier promotions users are equally likely to acquire an app that they are aware of as they are to acquire an app they haven't heard of, exactly because the promote is low barrier; they have little reason not to go through with it. But with high barrier promotions, users are more likely to do the effort required to redeem the coupon or pay the (reduced) price if they are aware of the app already. 
\par Hence, we see that high barrier promotions induce a `rich-get-richer' effect, where the increase an app observes during the promotion is proportional to their previous popularity. In contrast, apps participating in low barrier promotions receive, on average, the same increase no matter their previous popularity.

\paragraph{Age} The age of an app can be an indicator of how established it is and how much time it had to build its user base and word of mouth. We see in Table \ref{regressions_characteristics} that the coefficient for the age variable is not significantly different from zero for all of the studied promotions except the Apple Amazing one, the only promotion that offers partial discount. This could indicate the users are more willing to pay even a reduced price for apps that have been in the market longer, and hence, perhaps, had the time to release multiple versions fixing any bugs and improving their product. In contrast, in free promotions the app's success is not dependent on it's age.

\paragraph{Size} Finally, we use the size of the offered app as an indicator for the level of sophistication of the source code as well as characteristics like graphics, number of gameplay levels (if the app is a game) and more. We see in Table \ref{regressions_characteristics}, that the coefficient for the size variable is not significantly different from zero for any of the studied promotions. This means that when promoted, apps of all sizes benefit, on average, the same from the promotion.
\par Our insights from this section can help practitioners better design their own promotions according to their apps. For examplee, apps that are newer and not yet very succesfull should avoid being featured in high barrier promotions alongside with more successful apps, since the more successful apps will disproportionally benefit from the promotion.

\section{Effect of the Promotions on the Competition}
\par Many online stores today use machine learning algorithms that, based on user's behavior on the site try to recommend items that the user might be interested in. For example, Amazon has a section called `Customers Who Bought This Item Also Bought' on the product pages of their item.  Such recommendation algorithms can help users explore and discover new products of interest in an otherwise very large search space of options (e.g., Amazon and eBay have millions of products each for sale, Google's and Apple's App Store have more than 1.5 million apps each).
\par Apple's recommendation engine in the AppStore works mainly by displaying a `Customers also bough' tab on the product page of many apps. Even though sometimes these apps can be complimentary (such as Facebook and Messenger) most of the times they seem to be competing. For example, a casual game will have recommendations for other casual games and a productivity app will have recommendations for other apps with similar functions. These similar apps, compete for the user's attention, wallet as well as phone storage space (the 16GB model is the best selling amongst the iPhones).
\par For every promoted app we gathered the apps appearing in its `Customers also bought' tab'. We then collected the entire review history of every such app. We aim to study the effect of the promotions on these competing apps' sales. A negative effect on sales would suggest that the promotion causes customers to select the promoted app amongst the alternatives, possibly because of it's discounted price or to avoid the cost of searching further. A positive effect would suggest that the promoted app increases awareness for itself as well as apps similar to it and customers use the promoted app as a {\it starting point} to search further for the best alternative. 

\subsection{Econometric Model}
\par In order to statistically test these findings, we estimate the following model on the volume of reviews for the apps similar to the four set of featured apps. 

\begin{equation}\label{model_1}
\mathrm{sales}= \alpha_0 + \beta_1\mathrm{post} +\beta_2\mathrm{treat} + \beta_3\mathrm{post}\cdot\mathrm{treat}+e
\end{equation}

The post variable indicates if the datapoint is after the promotion, and treat is a binary variable indicating if the review is for an app in the treatment group or not. The post$\cdot$treat interaction term is the one that captures the variation that is not explained by the control dataset, hence it's the variable that estimates the causal effect of the promotion. Note that, unlike previous sections, the treatment group here is not the apps featured in the promotions but the apps that appear in the `Customers also bought' section of the the promoted apps.

\par Table \ref{regressions_comp} summarizes the results of the estimations of Model \ref{model_1} for the four studied promotions.

\begin{table*}[h!]
\begin{center}
%\textsc{}\\
\vspace{0.5cm}
\begin{tabular}{l c c c c  c} 
 \toprule
		{\it \small Promotion}				&Intercept 		&post 		& treat  	&  $\mathrm{post}\cdot\mathrm{treat}$\\[6pt]% & R-squared \\[6pt]
\midrule
\multirow{2}{*}{Apple Weekly} 		& 5.70			&-0.71			&$62.65^{***}$			& $-23.66^{**}$	\\[6pt]	%		&0.04			\\[6pt]
									&(3.19)				&(4.47)				&(5.02)			&(6.85)				\\[6pt]

\midrule

\multirow{2}{*}{\small Starbucks Digital} &   $5.65^{***}$		& $1.73$	 	 &$15.22^{***}$ 		& $-6.97^{*}$\\[6pt]	%		&0.13			\\[6pt]
										&(1.34)				&(1.88)				&(2.27)				&(3.16)		\\[6pt]

\midrule

\multirow{2}{*}{Starbucks In-Store} 	& $5.65^{***}$		& 	$1.73$			&$18.73^{***}$ 		& $-8.25^{**}$	\\[6pt]%		&0.04		\\[6pt]
									&(1.15)					&(1.59)			&(1.98)				&(2.77)		\\[6pt]
\midrule

\multirow{2}{*}{Apple Amazing} 					&  $5.73^{***}$			& $-0.14$			&$11.523^{***}$ 		& $3.43^{**}$	\\[6pt]%	&0.003			\\[6pt]
									&(0.62)					&(0.85)			&(0.94)				&(1.28) &	\\[6pt]%	&\\[6pt]

%\midrule
%\multirow{2}{*}{Number of Tags} 	& $0.4500^{***}$	& $ -0.0526^{***}$		&$0.0008$ 		& $0.0002$			&0			\\[6pt]
	%								&(0.008)			&(0.05)					&(0.001)		&(0.001)		&\\[6pt]

\bottomrule
\end{tabular}
\end{center}
Values in parentheses are standard errors.\\
\small{$^{*}:p<0.05$, $^{**}: p<0.01$, $^{***}:p<0.001$}
\caption{The effect of the promotions on the volume of reviews of their competitors}
\label{regressions_comp}
\end{table*}

\par The post$\cdot$treat coefficients are the ones estimating the extend of the effect that cannot be explained by the control dataset.
We see that the coefficients are negative and significant for Apple Weekly and Starbucks In-Store, negative and barely not significant ($p$=0.08) for Starbucks Digital and positive and significant for Apple Amazing.

\subsection{Results}
\par Since low barrier promotions make it very easy and costless for the user to acquire the promoted app, we expect that customers exposed to such suers will not explore for further alternatives and hence competing sales will decrease. Indeed, we find that apps competing with Apple Weekly promotions see their average daily number of reviews fall by XX in the 7 days of the promotion compared to the 7 days before. Similarly, we see a decrease of YY for apps competing with Starbucks Digital apps. Furthermore, we find that even apps competing with the Starbucks In-Store apps see a decrease of ZZ. This is even though Starbucks In-Store has a slightly non-trivial redemption procedure. Finally, and perhaps most interestingly, we find an {\it increase} of ZZ for the sales of the apps competing with Apple Amazing apps. 
\par The negative and significant coefficients for the post$\cdot$treat variable for all full price discounted promotions, shown in Table \ref{regressions_comp}, confirm that such promotions have a negative effect on the sales of their competition. In contrast, the only partially discounted promotion causes an increase in the sales of its competition. 
\par This suggests that, especially in electronic markets where such recommendation algorithms are in place, increasing awareness for a product may increase awareness and sales for it's competition. That is, unless the promotion is such that offers no incentives for users to explore for further alternatives, such as a full price discount.
\par We expect similar positive and negative externalities to exists in other types of markets as well, and in other types of marketing campaigns. Essentially, if we look at the recommendation graph produced by the centralized platform as a graph on which the user is walking a random walk, any increase in the probability of a specific point will have some effects on that point's neighbors. 
\par This suggests that practitioners designing marketing campaigns for their products should be aware of the positive externalities they can have on their competition. In fact, phenomena similar to this has already been observed in the literature, \cite{lewis2014samsung}

\section{Discussion and Insights}
\par We studied four promotions offered on Apple's iOS AppStore, that varied in scale of exposure, level of price discount and ease of redemption. We aim to understand better the effects that these promotions have on the sales as well as the ratings of the featured apps. We found positive effects on the sales for all promotions and mixed effects on the ratings. Notably, the only promotion that was not full price discounted was that only one that had a negative effect on the ratings of its featured apps whereas the weekly promotion run by Apple was the one with the largest positive effect.
\par We also explored the correlation between app characteristics and how succesfull their promotion campaign is. We found that when a promotion is low barrier, i.e., digital and free, users are equally likely to get an app no matter the app's previous popularity, regular price and age. In contrast, promotions with higher barriers, such as non-zero price or non-trivial redemption process, induce a rich-get-effect where users are more likely to get a promoted app if it's already established and successful.
\par Finally, our study also explored the effect of the promotions on the sales of their competitors. We found that full price discounts decrease the sales of their competitors, perhaps because users have no incentive to search for further alternatives. In contrast, the only partially discounted promotion in our study, Apple Amazing, caused an increase in the sales of it's competition. This suggests that the promotion raised awareness for the featured apps, but the non-free price tag incentivized users to hold back on their purchase until they explore further alternatives. 
\par These insights can help practitioners better design their promotion campaigns. Our work displays the benefits in sales and ratings of being featured in carefully selected and far-reaching promotions, but also the potential risks of not full price discounts. We also show that apps are newer and not yet very popular should avoid being featured in high barrier promotions alongside with more successful apps, since then the more succesfull apps will disproportionally benefit from the promotion. Finally, we exhibit the positive and negative externalities that various types of promotions can have, which make developers more aware of the full extend of the effect that their promotions can have on their competitors, as well as the other way around too.

\section{Future Directions}

%\par Not surprisingly, being featured on high profile marketing campaigns and promotions helps increase sales for mobile apps. More interestingly, this increase in sales does not have to come with a decrease on ratings especially if the promoted apps are chosen with the proper incentives or some simple mechanisms are put in place to filter out customers that are not part of the target group of the app. Such mechanisms can be low but not-zero prices and slightly non-trivial redemption procedures.

%\par Large scale promotions not only help increase the sales of the promoted apps but also help decrease the sale of their competition, as evidenced by the three of the four sets of promoted apps studied in this paper. It seems crucial though that the reduced price is lower than the alternatives' otherwise, as evidenced by the Apple Amazing promotion, the competition can benefit from and steal some of the spotlight the promoted apps receives. 
\par We believe there are many interesting questions left to be studied in future work. First, most of the other major app store platforms have their own variation of a `Free App of the Week/Day' promotion. Investigating if similar results hold for the other app stores can increase our understanding on the subject and provide further insights for practitioners to better design their promotions on various platform. 
\par Furthermore, it's interesting to see what is the effect of promotions that are not accompanied by any price discount, such as the `Editor's Choice' list. These promotions signal that a high profile entity with knowledge on the matter (usually the App Store itself) is endorsing an app for its high quality. What is the effect of such signals on the ratings and sales of the featured apps?
\par Moreover, promotions done by the developer's themselves can also be of interest. The four promotions studied in this paper all involve a selection process by a third party, but at any given day there are apps that are being sold in reduced prices, from their developers. These promotions are closer in spirit to the ones offered on platforms like Groupon since they are self-selected. Insights on the effect that these type of promotions have on the sales and ratings of the apps can help our understanding of the `Groupon effect' \cite{byers2012groupon}. It can also help developers explore better promotion strategies.
\par Finally, it will be of interest to see if effects similar to the ones discovered in this study arise when other types of digital goods, such as songs and movies, are offered in promotion. 

\bibliographystyle{plainnat}
\bibliography{large_promotions} % if more than one, comma separated

\begin{thebibliography}{19}
\providecommand{\natexlab}[1]{#1}
\providecommand{\url}[1]{\texttt{#1}}
\expandafter\ifx\csname urlstyle\endcsname\relax
  \providecommand{\doi}[1]{doi: #1}\else
  \providecommand{\doi}{doi: \begingroup \urlstyle{rm}\Url}\fi

\bibitem[Adamopoulos and Todri(2014)]{adamopoulos2014social}
Panagiotis Adamopoulos and Vilma Todri.
\newblock Social media analytics: The effectiveness of promotional events on
  brand user base in social media.
\newblock 2014.

\bibitem[Ajorlou et~al.(2014)Ajorlou, Jadbabaie, and
  Kakhbod]{ajorlou2014dynamic}
Amir Ajorlou, Ali Jadbabaie, and Ali Kakhbod.
\newblock Dynamic pricing in social networks: The word of mouth effect.
\newblock \emph{Available at SSRN 2495509}, 2014.

\bibitem[Anderson and Magruder(2012)]{anderson2012learning}
Michael Anderson and Jeremy Magruder.
\newblock Learning from the crowd: Regression discontinuity estimates of the
  effects of an online review database*.
\newblock \emph{The Economic Journal}, 122\penalty0 (563):\penalty0 957--989,
  2012.

\bibitem[Basuroy et~al.(2003)Basuroy, Chatterjee, and
  Ravid]{basuroy2003critical}
Suman Basuroy, Subimal Chatterjee, and S~Abraham Ravid.
\newblock How critical are critical reviews? the box office effects of film
  critics, star power, and budgets.
\newblock \emph{Journal of Marketing}, 67\penalty0 (4):\penalty0 103--117,
  2003.

\bibitem[Byers et~al.(2012)Byers, Mitzenmacher, and Zervas]{byers2012groupon}
John~W Byers, Michael Mitzenmacher, and Georgios Zervas.
\newblock The groupon effect on yelp ratings: a root cause analysis.
\newblock In \emph{Proceedings of the 13th ACM Conference on Electronic
  Commerce}, pages 248--265. ACM, 2012.

\bibitem[Carare(2012)]{carare2012impact}
Octavian Carare.
\newblock The impact of bestseller rank on demand: Evidence from the app
  market*.
\newblock \emph{International Economic Review}, 53\penalty0 (3):\penalty0
  717--742, 2012.

\bibitem[Chevalier and Mayzlin(2006)]{chevalier2006effect}
Judith~A Chevalier and Dina Mayzlin.
\newblock The effect of word of mouth on sales: Online book reviews.
\newblock \emph{Journal of marketing research}, 43\penalty0 (3):\penalty0
  345--354, 2006.

\bibitem[Chintagunta et~al.(2010)Chintagunta, Gopinath, and
  Venkataraman]{chintagunta2010effects}
Pradeep~K Chintagunta, Shyam Gopinath, and Sriram Venkataraman.
\newblock The effects of online user reviews on movie box office performance:
  Accounting for sequential rollout and aggregation across local markets.
\newblock \emph{Marketing Science}, 29\penalty0 (5):\penalty0 944--957, 2010.

\bibitem[Cui et~al.(2012)Cui, Lui, and Guo]{cui2012effect}
Geng Cui, Hon-Kwong Lui, and Xiaoning Guo.
\newblock The effect of online consumer reviews on new product sales.
\newblock \emph{International Journal of Electronic Commerce}, 17\penalty0
  (1):\penalty0 39--58, 2012.

\bibitem[Dellarocas et~al.(2005)Dellarocas, Awad, and
  Zhang]{dellarocas2005using}
Chrysanthos Dellarocas, Neveen Awad, and M~Zhang.
\newblock Using online ratings as a proxy of word-of-mouth in motion picture
  revenue forecasting.
\newblock Technical report, Citeseer, 2005.

\bibitem[Duan et~al.(2008)Duan, Gu, and Whinston]{duan2008dynamics}
Wenjing Duan, Bin Gu, and Andrew~B Whinston.
\newblock The dynamics of online word-of-mouth and product sales?an empirical
  investigation of the movie industry.
\newblock \emph{Journal of retailing}, 84\penalty0 (2):\penalty0 233--242,
  2008.

\bibitem[Edelman et~al.(2011)Edelman, Jaffe, and Kominers]{edelman2011groupon}
Benjamin Edelman, Sonia Jaffe, and Scott~Duke Kominers.
\newblock To groupon or not to groupon: The profitability of deep discounts.
\newblock \emph{Marketing Letters}, pages 1--15, 2011.

\bibitem[Engstrom and Forsell(2014)]{engstrom2014demand}
Per Engstrom and Eskil Forsell.
\newblock Demand effects of consumers' stated and revealed preferences.
\newblock \emph{Available at SSRN 2253859}, 2014.

\bibitem[Lewis and Nguyen(2014)]{lewis2014samsung}
Randall~A Lewis and Dan~Tri Nguyen.
\newblock A samsung ad for the ipad? display advertising's competitive
  spillovers to search.
\newblock \emph{Display Advertising's Competitive Spillovers to Search (January
  2, 2014)}, 2014.

\bibitem[Liu(2006)]{liu2006word}
Yong Liu.
\newblock Word of mouth for movies: Its dynamics and impact on box office
  revenue.
\newblock \emph{Journal of marketing}, 70\penalty0 (3):\penalty0 74--89, 2006.

\bibitem[Luca(2011)]{luca2011reviews}
Michael Luca.
\newblock Reviews, reputation, and revenue: The case of yelp. com.
\newblock Technical report, Harvard Business School, 2011.

\bibitem[Reinstein and Snyder(2005)]{reinstein2005influence}
David~A Reinstein and Christopher~M Snyder.
\newblock The influence of expert reviews on consumer demand for experience
  goods: A case study of movie critics*.
\newblock \emph{The journal of industrial economics}, 53\penalty0 (1):\penalty0
  27--51, 2005.

\bibitem[Spriensma(2012)]{spriensma2012impact}
Gert~Jan Spriensma.
\newblock The impact of app discounts and the impact of being a featured app.
\newblock \emph{Distimo Publicatio}, 2012.

\bibitem[Zhu and Zhang(2010)]{zhu2010impact}
Feng Zhu and Xiaoquan Zhang.
\newblock Impact of online consumer reviews on sales: The moderating role of
  product and consumer characteristics.
\newblock \emph{Journal of Marketing}, 74\penalty0 (2):\penalty0 133--148,
  2010.

\end{thebibliography}

%\bibliographystyle{abbrv}
%\bibliography{large_promotions}
%\bibliographystyle{plain}

%\input{appendix}

\end{document}